\documentclass[
    ,final            
  ]
  {aipproc}

\layoutstyle{6x9}

\newcommand{\bx}{{\bf x}}

\newcommand{\lb}{\label}
\newcommand{\be}{\begin{equation}}
\newcommand{\ee}{\end{equation}}
\newcommand{\ber}{\begin{eqnarray}}
\newcommand{\eer}{\end{eqnarray}}
\newcommand{\bers}{\begin{eqnarray*}}
\newcommand{\eers}{\end{eqnarray*}}

\newcommand{\Bell}{\hbox{\boldmath $\ell$}}

\begin{document}

\title{Heat Transfer and Reconnection Diffusion in Turbulent Magnetized Plasmas }


\author{A. Lazarian}{
  address={Department of Astronomy, University of Wisconsin-Madison, lazarian@astro.wisc.edu}
}

\begin{abstract}
It is well known that magnetic fields constrain motions of charged particles, impeding the diffusion of charged particles perpendicular to magnetic field direction. This modification of transport processes is of vital importance for a wide variety of astrophysical processes including cosmic ray transport, transfer of heavy elements in the interstellar medium, star formation etc. Dealing with these processes one should keep in mind that, in realistic astrophysical conditions, magnetized fluids are turbulent. In this review we single out a particular transport process, namely, heat transfer and consider how it occurs in the presence of the magnetized turbulence. We show that the  ability of magnetic field lines to constantly change topology and connectivity is at the heart of the correct description of the 3D magnetic field stochasticity in turbulent fluids. This ability is ensured by fast magnetic reconnection in turbulent fluids and puts forward the concept of reconnection diffusion at the core of the physical picture of heat transfer in astrophysical plasmas. Appealing to reconnection diffusion we describe the ability of plasma to diffuse between different magnetized eddies explaining the advection of the heat by turbulence. Adopting the structure of magnetic field that follows from the modern understanding of MHD turbulence, we also discuss thermal conductivity that arises as electrons stream along stochastic magnetic field lines. We compare the effective heat transport that arise from the two processes and conclude that, in many astrophysically-motivated cases, eddy advection of heat dominates. Finally, we discuss the concepts of sub and superdiffusion and show that the subdiffusion requires rather restrictive settings. At the same time, accelerated diffusion or superdiffusion of heat perpendicular to the mean magnetic field direction is possible on the scales less than the injection scale of the turbulence.
\end{abstract}

\maketitle


\section{Main idea and structure of the review}

~~Heat transfer in turbulent magnetized plasma is an important astrophysical
problem which is relevant to the wide variety of circumstancies from mixing layers
in the Local Bubble (see Smith \& Cox 2001) and Milky way
(Begelman \& Fabian 1990) to cooling flows in intracluster medium (ICM)
(Fabian 1994). The latter problem has been subjected to particular scrutiny
as  observations
do not support the evidence for the cool gas (see Fabian et al. 2001).
This is suggestive of the existence of heating that replenishes the
energy lost via X-ray emission. Heat transfer from hot outer regions is
an important process to consider in this context.

~~It is well known that magnetic fields can suppress thermal conduction perpendicular to their direction. However,
this is true for laminar magnetic field, while astrophysical plasmas are generically turbulent (see Armstrong et al 1994, Chepurnov \& Lazarian 2010). The issue of heat transfer in realistic turbulent magnetic fields has been long debated (see Bakunin 2005 and references therein).

~~Below we argue that turbulence changes the very nature of the process of heat transfer. To understand the differences between laminar and turbulent cases one should consider both motion of charged particles along turbulent magnetic fields and turbulent motions of magnetized plasma that also transfer heat. The description of both processes require the knowledge of the dynamics of magnetic field lines and the structure of the magnetic field lines in turbulent flows. The answers to these questions are provided by the theories of magnetic reconnection and magnetic turbulence. To provide the quantitative estimates of the heat transfer the review addresses both theories, discussing the generic process of reconnection diffusion which describes the diffusion induced by the action of turbulent motions in the presence of reconnection. We stress the fundamental nature of the process which apart from
heat transfer is also important e.g. for removing magnetic field in star formation process (Lazarian 2005).

~~In \S 2 we discuss the omnipresence of turbulence in astrophysical fluids, introduce major ideas of MHD turbulence theory and turbulent magnetic reconnection in \S 3 and \S 4, respectively, relate the concept of reconnection diffusion to the processes of heat transfer in magnetized plasmas in \S5. We provide detailed discussion of heat conductivity via streaming electrons in \S 6, consider heat advection by turbulent eddies in \S 7, and compare the efficiencies of the latter two processes in \S 8. Finally, we discuss heat transfer on scales smaller than the turbulence injection scale in \S 9 and provide final remarks in \S 10.

\section{Magnetized turbulent astrophysical media}

~~Astrophysical plasmas are known to be magnetized and turbulent. Magnetization of these fluids most frequently arises from the dynamo action to which turbulence is an essential component (see Schekochihin et al. 2007). In fact, it has been shown that turbulence in weakly magnetized conducting fluid converts about ten percent of the energy of the cascade into the magnetic field (see Cho et al. 2009). This fraction does not depend on the original magnetization and therefore magnetic fields will come to equipartition with the turbulent motions in about 10 eddy turnover times.

~~We deal with magnetohydrodynamic (MHD) turbulence which provides a correct fluid-type description of plasma turbulence at large scales\footnote{It is possible to show that in terms magnetic field wandering that is important, as we see below, for heat
transfer the MHD description is valid in collisionless regime of magnetized plasmas (Eyink, Lazarian \& Vishniac (2011).}. Astrophysical turbulence is a direct consequence of large scale fluid motions experiencing low friction. This quantity is described by Reynolds number $Re\equiv LV/\nu$, where $L$ is the scale of fluid motions, $V$ is the velocity at this scale and $\nu$ is fluid viscosity. The Reynolds numbers are typically very large in astrophysical flows as the scales are large. As magnetic fields decrease the viscosity for the plasma motion perpendicular to their direction, $Re$ numbers get really astronomically large. For instance, $Re$ numbers of $10^{10}$ are very common for astrophysical flow. For so large $Re$ the inner degrees of fluid motion get excited and a complex pattern of motion develops.

~~The drivers of turbulence, e.g. supernovae explosions in the interstellar medium, inject energy at large scales and then the energy cascades down to small scales through a hierarchy of eddies spanning up over the entire inertial range. The famous Kolmogorov picture (Kolmogorov 1941) corresponds to hydrodynamic turbulence, but, as we discuss further, a qualitatively similar turbulence also develops in magnetized fluids/plasmas.

~~Simulations of interstellar medium, accretion disks and other astrophysical environments also produce turbulent picture, provided that the simulations are not dominated by numerical viscosity. The latter requirement is, as we see below, is very important for the correct reproduction of the astrophysical reality with computers.

~~The definitive confirmation of turbulence presence comes from observations, e.g. observations of electron density fluctuations in the interstellar medium, which produce a so-called Big Power Law in the Sky (Armstrong et al. 1994, Chepurnov \& Lazarian 2010), with the spectral index coinciding with the Kolmogorov one. A more direct piece of evidence comes from the observations of spectral lines. Apart from showing non-thermal Doppler broadening, they also reveal spectra of supersonic turbulent velocity fluctuations when analyzed with techniques like Velocity Channel Analysis (VCA) of Velocity Coordinate Spectrum (VCS) developed (see Lazarian \& Pogosyan 2000, 2004, 2006, 2008) and applied to the observational data (see Padoan et al. 2004, 2009, Chepurnov et al. 2010) rather recently.

~~All in all, the discussion above was aimed at conveying the message that the turbulent state of magnetized astrophysical fluids is a rule and therefore the discussion of any properties of astrophysical systems should take this state into account. We shall show below that both magnetic reconnection and heat transfer in magnetized fluids are radically changed by turbulence.

\section{Strong and weak Alfvenic turbulence}

~~For the purposes of heat transfer, Alfvenic perturbations are most important. Numerical studies in Cho \& Lazarian (2002, 2003) showed that the Alfvenic turbulence develops an independent cascade which is marginally affected
by the fluid compressibility. This observation corresponds to theoretical expectations of the Goldreich \& Sridhar (1995) theory that we briefly describe below (see also Lithwick \& Goldreich 2001). In this respect we note that the MHD approximation is widely used to describe the actual magnetized plasma turbulence over scales that are much larger than both the mean free path of the particles and their Larmor
radius (see Kulsrud 2004 and ref. therein). More generally, the most important incompressible Alfenic part of the plasma motions can described by MHD even below the mean free path (see Eyink et al. 2011 and ref. therein).

~~While having a long history of ideas, the theory of MHD turbulence has become
testable recently due to the advent numerical simulations (see Biskamp 2003)
which confirm (see Cho \& Lazarian 2005 and
ref. therein) the prediction of magnetized Alfv\'enic eddies
being elongated in the direction of magnetic field (see Shebalin, Matthaeus \&
Montgomery 1983, Higdon 1984) and provided results consistent with the
quantitative relations for the degree of eddy elongation obtained  in Goldreich \& Sridhar (1995, henceforth GS95).

~~The hydrodynamic counterpart of the MHD turbulence theory is the famous Kolmogorov theory of turbulence. In that theory, energy is injected at large scales, creating large eddies which correspond to large $Re$ numbers and therefore do not dissipate energy through viscosity\footnote{Reynolds number $Re\equiv LV/\nu=(V/L)/(\nu/L^2)$ which is the ratio of the eddy turnover rate $\tau^{-1}_{eddy}=V/L$ and the viscous dissipation rate $\tau_{dis}^{-1}=\eta/L^2$. Therefore large $Re$ correspond to negligible viscous dissipation of large eddies over the cascading time $\tau_{casc}$ which is equal to $\tau_{eddy}$ in Kolmogorov turbulence.} but transfer energy to smaller eddies. The process continues till the cascade reaches the eddies that are small enough to dissipate energy over an eddy turnover time. In the absence of compressibility the hydrodynamic cascade of energy is $\sim v^2_l/\tau_{casc, l} =const$, where $v_l$ is the velocity at the scale $l$ and the cascading time for the eddies of size $l$ is $\tau_{cask, l}\approx l/v_l$. From this the well known relation $v_l\sim l^{1/3}$ follows.

~~Modern MHD turbulence theory can also be understood in terms of eddies. However, in the presence of dynamically important magnetic field, eddies cannot be isotropic. Any motions bending magnetic field should induce a back-reaction and Alfven waves propagating along the magnetic field. At the same time, one can imagine eddies mixing magnetic field lines perpendicular to the direction of magnetic field. For the latter eddies the original Kolmogorov treatment is applicable resulting perpendicular motions scaling as $v_ll_{\bot}^{1/3}$, where $l_{\bot}$ denotes scales measured perpendicular to magnetic field and correspond to the perpendicular size of the eddy. These mixing motions induce Alfven waves which determine the parallel size of the magnetized eddy.  The key stone of the GS95 theory is {\it critical balance}, i.e. the equality of the eddy turnover time $l_{\bot}/v_l$ and the period of the corresponding Alfven wave $\sim l_{\|}/V_A$, where $l_{\|}$ is the parallel eddy scale and $V_A$ is the Alfven velocity. Making use of the earlier expression for $v_l$ one can easily obtain $l_{\|}\sim l_{\bot}^{2/3}$, which reflects the tendency of eddies to become more and more elongated as energy cascades to smaller scales.

~~While the arguments above are far from being rigorous they correctly reproduce the basic scalings of magnetized turbulence when the velocity equal to $V_A$ at the injection scale $L$. The most serious argument against the picture is the ability of eddies to perform mixing motions perpendicular to magnetic field. We shall address this issue in \S 3 but for now we just mention in passing that strongly non-linear turbulence does not usually allow the exact derivations. It is numerical experiments that proved the above scalings for incompressible MHD turbulence (Cho \& Vishniac 2000, Maron \& Goldreich 2001, Cho, Lazarian \& Vishniac 2002) and for the Alfvenic component of the compressible MHD turbulence (Cho \& Lazarian 2002, 2003, Kowal \& Lazarian 2010).

~~It is important to stress that the scales $l_{\bot}$ and $l_{\|}$ are measured in respect to the system of reference related to the direction of the local magnetic field "seen" by the eddy. This notion was not present in the original formulation of the GS95 theory and was added in Lazarian \& Vishniac (1999) (see also Cho \& Vishniac 2000, Maron \& Goldreich 2001, Cho et al. 2002). In terms of mixing motions that we mentioned above it is rather obvious that the free Kolmogorov-type mixing is possible only in respect to the local magnetic field of the eddy rather than the mean magnetic field of the flow.

~~GS95 theory assumes the isotropic injection of energy at scale
$L$ and the injection velocity equal to the Alfv\'en velocity in
the fluid $V_A$, i.e. the Alfv\'en Mach number
$M_A\equiv (\delta V/V_A)=1$.  This model can be easily generalized
for both $M_A<1$ and $M_A>1$ at the injection (see Lazarian \& Vishniac 1999 and Lazarian 2006, respectively).  Indeed, if $M_A>1$,
instead of the driving scale $L$ for  one can use another scale,
namely $l_A$, which is the scale at
which the turbulent velocity gets equal to $V_A$.  For $M_A\gg 1$
magnetic fields are not dynamically important at the largest scales and the
turbulence at those scales follows the isotropic
 Kolmogorov cascade $v_l\sim l^{1/3}$ over the
range of scales $[L, l_A]$. This provides $l_A\sim L M_A^{-3}$.
If $M_A<1$, the turbulence obeys GS95 scaling (also called ``strong''
MHD turbulence) not from the scale $L$, but from a smaller scale $l_{trans}
\sim L M_A^{2}$ (Lazarian \& Vishniac 1999), while in the range
$[L, l_{trans}]$ the turbulence is ``weak''.

~~The properties of weak and strong turbulence are rather different. The weak turbulence is wave-like turbulence with wave packets undergoing many collisions before transferring energy to small scales\footnote{Weak turbulence, unlike the strong one, allows an exact analytical treatment (Gaultier et al. 2002).}. On the contrary, the strong turbulence is eddy-like with cascading happening similar to Kolmogorov turbulence within roughly an eddy turnover time.  One also should remember that the notion "strong" should not be associated with the amplitude of turbulent motions, but only with the strength of the non-linear interaction. As the weak turbulence evolves, the interactions of wave packets increases as the ratio of the parallel to perpendicular scales of the packets increases making the turbulence strong. In this case, the amplitude of the perturbations may be very small.

~~While there ongoing debates whether the original GS95 theory should be modified to better describe MHD turbulence, we believe that, first of all, we do not have compelling evidence that GS95 is not adequate\footnote{Recent work by Beresnyak \& Lazarian (2010) shows that present day numerical simulations are unable to reveal the actual inertial range of MHD turbulence making the discussions of the discrepancies of the numerically measured spectrum and the GS95 predictions rather premature. In addition, new higher resolution simulations by Beresnyak (2011) reveal the predicted $-5/3$ spectral slope.}. Moreover, the proposed additions to the GS95 model do not change the nature of the physical processes that we present below.

~~The quantitative picture of astrophysical turbulence sketched in this section gives us a way to proceed with the quantitative description of key processes necessary to describe heat transfer. The interaction of fundamental MHD
modes within the cascade of compressible magnetized turbulence is described in Cho \& Lazarian (2005), but this
interaction is not so important for the processes of heat transfer that we discuss below.

\section{Magnetic reconnection of turbulent magnetic flux}

~~Magnetic reconnection is a fundamental process that violates magnetic flux being frozen in within highly conductive fluids. Intuitively one may expect that magnetic fields in turbulent fluids cannot be perfectly frozen in. Theory that we
describe below provide quantitative estimates of the violation of frozen in condition within turbulent fluids.

~~We would like to stress that the we are discussing the case of dynamically important magnetic field, including the case of weakly turbulent magnetic field. The case of weak magnetic field which can be easily stretched and bended by turbulence at any scale up to the dissipation one is rather trivial and of little astrophysical significance\footnote{In the case of dynamically unimportant field, the magnetic dissipation and reconnection happens on the scales of the Ohmic diffusion scale and the effects of magnetic field on the turbulent cascade are negligible. However, turbulent motions transfer an appreciable portion of the cascading energy into magnetic energy (see
Cho et al. 2010). As a result, the state of intensive turbulence with negligible magnetic field is short-lived.}. At the same time, at sufficiently small scales magnetic fields get dynamically important even for superAlfvenic turbulence.

\begin{figure}
\centering
  \includegraphics[height=.45\textheight]{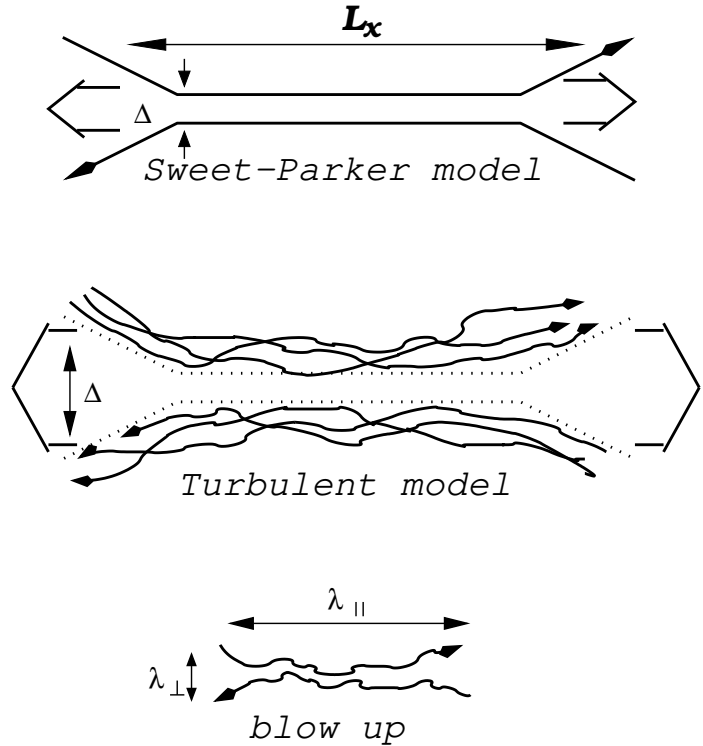}
  \caption{{\it Upper panel}: Sweet-Parker reconnection. $\Delta$ is limited by resistivity and small. {\it Middle panel}: reconnection according to LV99 model. $\Delta$ is determined by turbulent field wandering and can be large. {\it Lower panel}: magnetic field reconnect over small scales. From Lazarian, Vishniac \& Cho (2004).}
 \label{LV}
\end{figure}

~~Within the picture of eddies mixing perpendicular to the local magnetic field that we provided in the previous section, it is suggestive that magnetized eddies can provide turbulent advection of heat similar to the ordinary hydrodynamic eddies. This is rather counter-intuitive notion in view of the well-entrenched idea of flux being frozen in astrophysical fluids. As it is explained in Eyink et al. (2011) the frozen-in condition is not a good approximation for the turbulent fluids\footnote{Formal mathematical arguments on how and why the frozen-in condition fails may be found in Eyink (2011).}. The violation of the perfect frozenness of the magnetic field in plasmas also follows from LV99 model  of reconnection (see discussion in Vishniac \& Lazarian 1999).

~~A picture of two flux tubes of different directions which get into contact in 3D space is the generic framework to
describe magnetic reconnection. The upper panel of Figure \ref{LV} illustrates why reconnection is so slow in the textbook Sweet-Parker model. Indeed, the model considers magnetic fields that are laminar and therefore the frozen-in condition for magnetic field is violated only over a thin layer dominated by plasma resistivity. The scales over which the resistive diffusion is important are microscopic and therefore the layer is very thin, i.e. $\Delta\ll L_x$, where $L_x$ is the scale at which magnetic flux tubes come into contact. The latter is of the order of the diameter of the flux tubes and typically very large for astrophysical conditions. During the process of magnetic reconnection all the plasma and the shared magnetic flux\footnote{Figure \ref{LV}  presents only a cross section of the 3D reconnection layer. A shared component of magnetic field is going to be present in the generic 3D configurations of reconnecting magnetic flux tubes.} arriving over an astrophysical scale $L_x$ should be ejected through a microscopic slot of thickness $\Delta$. As the ejection velocity of magnetized plasmas is limited by Alfven velocity $V_A$, this automatically means that the velocity in the vertical direction, which is reconnection velocity, is much less than $V_A$.

~~The LV99 model generalizes the Sweet-Parker one by accounting for the existence of magnetic field line stochasticity (Figure \ref{LV} (lower panels)).  The depicted turbulence is sub-Alfvenic with relatively small fluctuations of the magnetic field. At the same time turbulence induces magnetic field wandering. This wandering was quantified in LV99 and it depends
on the intensity of turbulence. The vertical extend of wandering of magnetic field lines that at any point get into contact with the field of the other flux tube was identified in LV99 with the width of the outflow region. Note, that  magnetic field wandering is a characteristic feature of magnetized turbulence in 3D. Therefore, generically in
turbulent reconnection the outflow is no more constrained by the narrow resistive layer, but takes place through a much wider area $\Delta$ defined by wandering magnetic field lines. The extend of field wandering determines
the reconnection velocity in LV99 model.

~~An important consequence of the LV99 reconnection is that as turbulence
amplitude increases, the outflow region and therefore reconnection rate also increases, which entails the
ability of reconnection to change its rate depending on the level of turbulence. The latter is important both
for understanding the dynamics of magnetic field in turbulent flow and for explaining flaring reconnection events, e.g. solar flares.

~~We should note that the magnetic field wandering is mostly due to Alfvenic turbulence. To describe the field
wondering for weakly turbulent case LV99 extended the GS95 model for a subAlfvenic case. The same field wandering\footnote{As discussed in LV99 and in more details in Eyink et al. (2011) the magnetic field wandering, turbulence and magnetic reconnection are very tightly related concepts. Without magnetic reconnection, properties of magnetic turbulence and magnetic field wandering would be very different. For instance, in the absence of fast reconnection, the formation of magnetic knots arising if magnetic fields were not
able to reconnect would destroy the self-similar cascade of Alfvenic turbulence. The rates predicted by LV99 are
exactly the rates required to make Goldreich-Sridhar model of turbulence self-consistent.}, as we discuss later, is important for heat transfer by electrons streaming along magnetic field lines.

~~The predictions of the turbulent reconnection rates in LV99 were successfully tested 3D numerical simualtions in Kowal et al. (2009) (see also Lazarian et al. 2010 for an example of higher resolution runs). This testing provided  stimulated work on the theory applications,
e.g. its implication for heat transfer. One should keep in mind that the LV model assumes that the magnetic field flux tubes can come at arbitrary angle, which corresponds to the existence of shared or guide field within the reconnection layer\footnote{The model in LV99 is three dimensional and it is not clear to what extend it can be applied to 2D turbulence (see discussion in ELV11 and references therein). However, the cases of pure 2D reconnection and 2D turbulence are of little practical importance.}.

~~Alternative models of magnetic reconnection appeal to different physics to overcome the constraint of the Sweet-Parker model. In the Petcheck (1964) model of reconnection the reconnection layer opens up to enable the
outflow which thickness does not depend on resistivity. To realize this idea inhomogeneous resistivity, e.g. anomalous resisitivity associated with plasma effects, is required (see Shay \& Drake 1998). However, for turbulent plasmas, the effects arising from modifying the local reconnection events by introducing anomalous
resistivity are negligible as confirmed e.g. in Kowal et al. (2009).  Other effects, e.g. formation and ejection of plasmoids (see Shibata \& Tanuma 2001, Lorreiro et al. 2008) which may be important for initially laminar environments are not likely to play the dominant role in turbulent plasmas either. Therefore in what follows dealing
with turbulent transfer of hear we shall appeal to the LV99 model of reconnection.

\section{Reconnection diffusion and heat transfer}

\begin{figure}
  \includegraphics[height=.45\textheight]{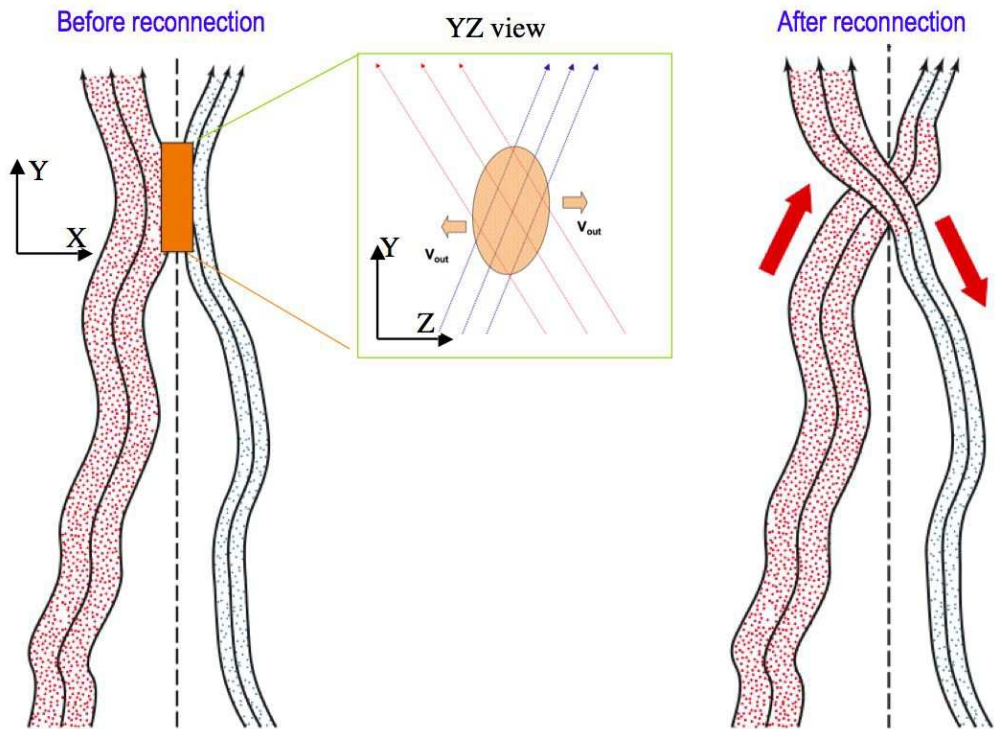}
  \caption{Diffusion of plasma in inhomogeneous magnetic field. 3D magnetic flux tubes get into contact and after reconnection plasma streams along magnetic field lines. {\it Right panel}: XY projection before reconnection, upper panel shows that the flux tubes are at angle in X-Z plane. {\it Left Panel}: after reconnection. }
  \label{recdiff}
\end{figure}

~~In the absence of the frozen-in condition in turbulent fluids one can talk about reconnection diffusion in magnetized turbulent astrophysical plasmas. The concept of reconnection diffusion is based on LV99 model and was first discussed in Lazarian (2005) in terms of star formation\footnote{Indeed, the issue of flux being conserved within the cloud presents a problem for collapse of clouds with strong magnetic field. These clouds also called subcritical were believed to evolve with the rates determined by the relative drift of neutrals and ions, i.e. the ambipolar diffusion rate.}. However, reconnection diffusion is a much broader concept applicable to different astrophysical
processes, including heat transfer in magnetized plasmas. In what follows we shall discuss several processes that enable heat transfer perpendicular to the mean magnetic field in the flow.

~~The picture frequently presented in textbooks may be rather misleading. Indeed, it is widely assumed that
magnetic field lines always preserve their identify in highly conductive plasmas even in turbulent flows. In this
situation the diffusion of charged particles perpendicular to magnetic field lines is very restricted. For instance,
the mass loading of magnetic field lines does not change to a high degree and density and magnetic field compressions follow each other. All these assumptions are violated in the presence of reconnection diffusion.

~~We shall first illustrate the reconnection diffusion process showing how it allows plasma to move perpendicular to the mean inhomogeneous magnetic field (see Figure \ref{recdiff}). Magnetic flux tubes with entrained plasmas intersect each other at an angle and due to reconnection the identity of magnetic field lines change. Before the reconnection plasma pressure $P_{plasma}$ in the tubes is different, but the total pressure $P_{plasma}+P_{magn}$ is the same for two tubes. After reconnection takes place, plasma streams along newly formed magnetic field lines to equalize the pressure along two new flux tubes. The diffusion of plasmas and magnetic field takes place. The effect of this process is to make magnetic field and plasmas more homogeneously distributed in the absence of the external fields\footnote{If this process acts in the presence of gravity, as this is the case of star formation, the heavy fluid (plasma) will tend to get to the gravitating center changing the mass to flux ratio, which is important to star formation processes. In other words, reconnection diffusion can do the job that is usually associated with the action of ambipolar diffusion (see numerical simulations in Santos de Lima et al. (2010).}. In terms of heat transfer, the process mixes up plasma at different temperatures
if the temperatures of plasma volumes along different magnetic flux tubes were different.

~~If turbulence had only one scale of motions its action illustrated by Figure \ref{recdiff} would create every flux tube
columns of hot and cold gas exchanging heat with each other through the diffusion of charged particles along
magnetic field lines. This is not the case, however, for a turbulence with an extended inertial cascade. Such a turbulence would induce mixing depicted in Figure \ref{recdiff} on every scale, mixing plasma at smaller and
smaller scales.

\begin{figure}
\centering
  \includegraphics[height=.45\textheight]{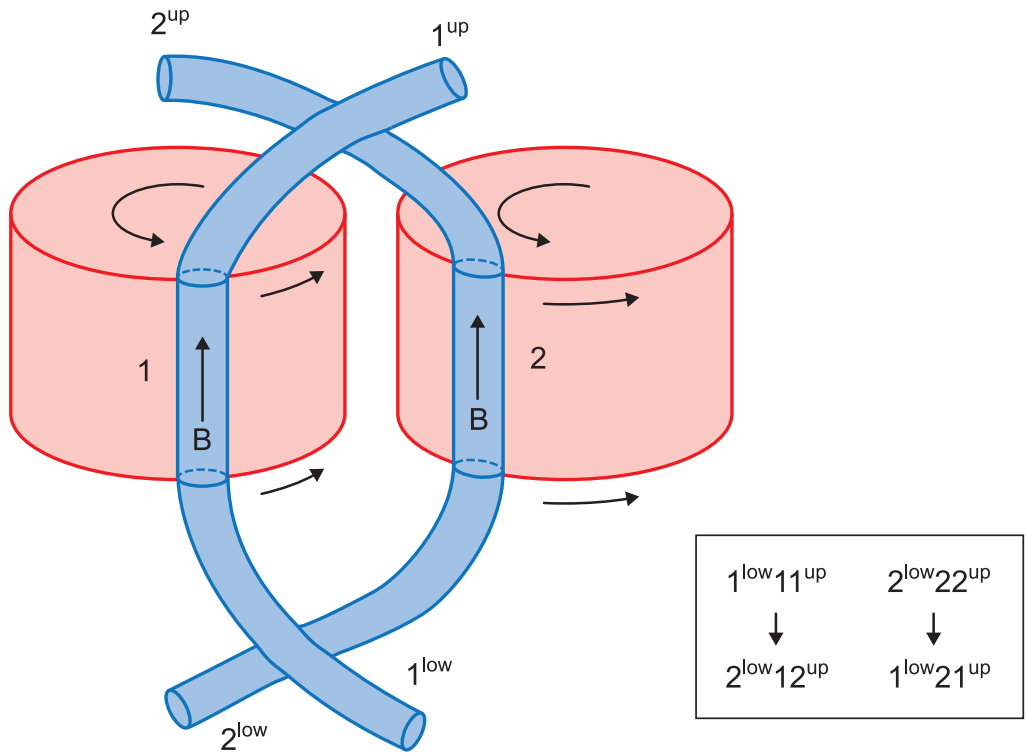}
  \caption{Exchange of plasma between magnetic eddies. Eddies carrying magnetic flux tubes interact through reconnection of the magnetic field lines belonging to two different eddies. This enables the exchange of matter between eddies and induces a sort of turbulent diffusivity of matter and magnetic field.}
  \label{mix}
\end{figure}

~~When plasma pressure along magnetic field flux tubes is the same, the connection of flux tubes which takes place in turbulent media as shown in Figure \ref{mix} is still important for heat transfer. The reconnected flux tubes illustrate the formation of the wandering magnetic field lines along which electron and ions can diffuse transporting heat. For the sake of simplicity, we shall assume that electrons and ions have the same temperature. In this situation, the transfer of heat by ions is negligible and for the rest of the presentation we shall talk about the transport of heat by electrons moving along wandering field lines\footnote{This is true provided that the current of diffusing hot electrons is compensated by the current of oppositely moving cold electrons, the diffusivity of electrons along wandering magnetic field lines is dominant compared with the diffusivity and heat transfer by protons and heavier ions. If there is no compensating current, electrons and ions are coupled by electric field and have to diffuse along wandering magnetic fields together and at the same rate. This could be the case of diffusion of plasmas into neutral gas. However, we do not discuss these complications here}.

~~Consider the above process of reconnection diffusion in more detail. The eddies 1 and 2 interact through the reconnection of the magnetic flux tubes associated with eddies.  LV99 model shows that in turbulent flows reconnection happens within one eddy turnover time, thus ensuring that magnetic field does not prevent free mixing motions of fluid perpendicular to the local direction of magnetic field. As a result of reconnection, the tube $1^{low} 1 1^{up}$ transforms into $2^{low}1 2^{up}$ and a tube $2^{low} 2 2^{up}$ transforms into $1^{low} 2 1^{up}$. If eddy 1 was associated with hotter plasmas and eddy 2 with colder plasmas, then the newly formed magnetic flux tubes will have both patches of hot and cold plasmas. For the hierarchy of eddies the shedding of entrained plasmas into hot and cold patches along the same magnetic field lines allows electron conductivity to remove the gradients, conducting heat. This is the process of turbulent advection of heat in magnetized plasmas.

~~The difference between the processes depicted in Figures \ref{recdiff} and \ref{mix} is due to the fact that
the process in Figure \ref{recdiff} is limited by the thermal velocity of particles, while the process in Figure \ref{mix}
depends upon the velocity of turbulent eddies only. In actual plasmas in the presence of temperature gradients
plasmas along different elementary flux tubes will have different temperature and therefore two processes will
take place simultaneously.

\begin{figure}
\centering
  \includegraphics[height=.45\textheight]{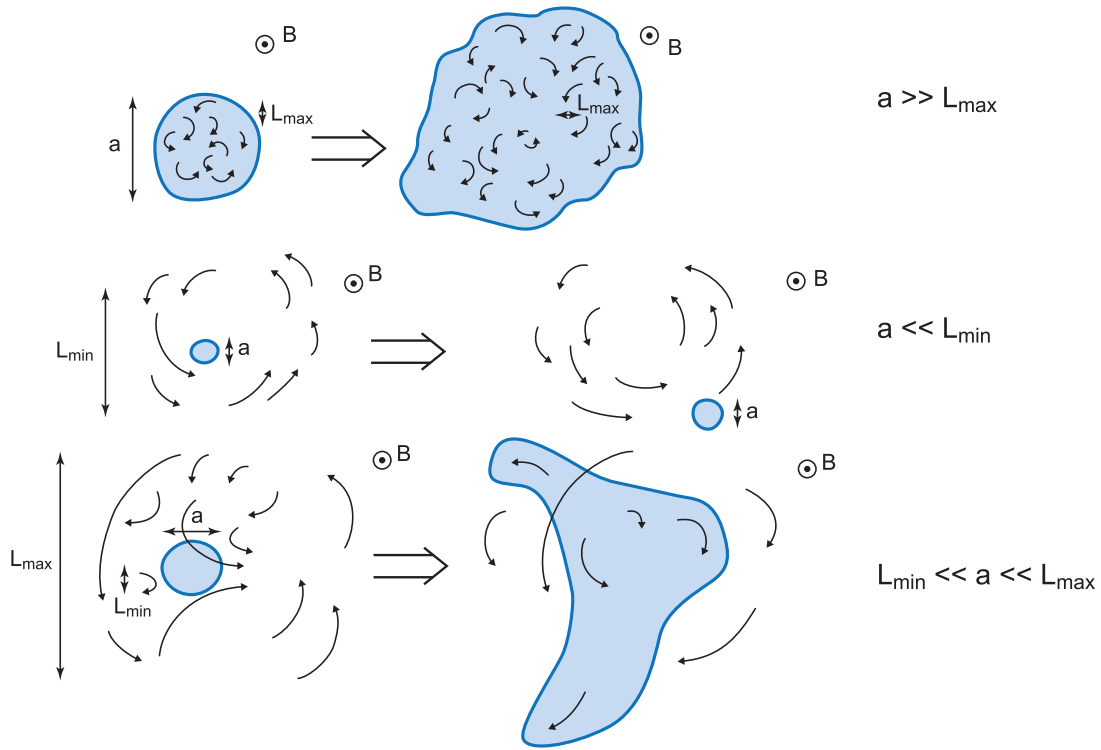}
  \caption{Heat diffusion depends on the scale of the hot spot. Different regimes emerge depending on the relation
  of the hot spot to the sizes of maximal and minimal eddies present in the turbulence cascade. Mean magnetic field 
  $B$ is directed perpendicular to the plane of the drawing. Eddies perpendicular to magnetic field lines correspond to Alfvenic turbulence. The plots illustrate heat diffusion for different regimes. {\it Upper} plot corresponds to the heat spot being less than the minimal size of turbulent eddies; {\it Middle plot} corresponds to the heat spot being less than the damping scale of turbulence;  {\it Lower plot} corresponds to the heat spot size within the inertial range of turbulent motions.}
  \label{turb}
\end{figure}

~~Whether the motion of electrons along wandering magnetic field lines or the dynamical mixing induced by turbulence is more important depends on the ratio of eddy velocity to the sonic one, the ratio of the turbulent motion scale to the mean free path of electrons and the degree of plasma magnetization. Strong magnetization both limits the efficiency of turbulent mixing perpendicular to magnetic field lines and the extent to which plasma streaming along magnetic field lines moves perpendicular to the direction of the mean field. However, but reduction of heat transfer efficiency is different for the two processes. We provide quantitative treatment of these processes in the next section.

~~An interesting example of practical interest is related to the diffusion of heat from a hot spot. This case of reconnection diffusion is illustrated by Figure \ref{turb}. In this situation heat transfer depends on whether
the scale of turbulent motions is larger or smaller than the hot spot. Consider this situation in more detail.
 Turbulence is characterized by its injection scale $L_{max}$, its dissipation scale $L_{min}$ and its inertial range $[L_{min}, L_{max}]$. The heat transfer depends on what scales we consider the process. Figure \ref{turb} illustrates our point.  Consider a hot spot of the size $a$ in turbulent flow and consider Alfvenic eddies perpendicular to magnetic field lines. If turbulent eddies are much smaller than
$a$, which is the case when $a\gg L_{min}$ they extend the hot spot acting in a random walk fashion. For eddies much larger than the hot spot, i.e. $a\ll L_{min}$  they mostly advect hot spot. If $a$ is the within the inertial range of turbulent motions, i.e. $L_{min}<a<L_{max}$ then a more complex dynamics of turbulent motions is involved. This is also the case where the field wandering arising from these motions is the most complex. Turbulent motions with the scale comparable with the hot spot induce a process of the accelerated Richardson diffusion (see more in \S 10).

~~In terms of practical simulation of reconnection diffusion effects, it is important to keep in mind that the LV99 model
predicts that the largest eddies are the most important for providing outflow in the reconnection zone and therefore the reconnection will not be substantially changed if turbulence does not have an extended inertial range. In addition, LV99 predicts that the effects of anomalous resistivity arising from finite numerical grids do not change the rate of turbulent reconnection. We note that both effects were successfully tested in Kowal et al. (2009).

\section{Heat conduction through streaming of electrons}

\subsection{General considerations}

~~As magnetic reconnection was considered by many authors even more mysterious than the heat transfer in plasmas, it is not surprising that the advection of heat by turbulent eddies was not widely discussed. Instead  for many year the researchers preferred to consider heat transfer by plasma conductivity along turbulent magnetic field lines (see Chandran \& Cowley 1998, Malyshkin \& Kulsrud 2001). This conductivity is mostly due to electrons streaming along magnetic field lines. Turbulent magnetic field lines allow streaming electrons to diffuse perpendicular to the mean magnetic field and spread due to the magnetic field wandering that we discussed earlier. Therefore the description of magnetic field wandering obtained in LV99 is also applicable for describing the processes of heat transfer.

~~We start with the case of trans-Alfvenic turbulence considered by Narayan \& Medvedev (2001, henceforth NM01).
They appeal to magnetic field wandering and obtained estimates of thermal
conductivity by electrons for the special case of turbulence velocity
$V_L$ at the energy
injection scale $L$ that is equal to the Alfven velocity $V_A$. As we discussed earlier
this special case is described by the original GS95 model and the Alfven Mach number $M_A\equiv (V_L/V_A)=1$.
We note that this case is rather
restrictive, as the intracuster medium (ICM) is superAlfvenic, i.e. $M_A>1$, while other
astrophysical situations, e.g. solar atmosphere, are subAlfvenic, i.e. $M_A<1$.  Different phases of
interstellar medium (ISM) (see Draine \& Lazarian 1998 and Yan, Lazarian \& Draine 2004 for lists of idealized ISM phases) present the cases of both superAlfvenic and subAlfvenic turbulence.

~~As we discussed above, the generalization of GS95 model of turbulence for subAlfvenic case is provided in LV99. This was employed in Lazarian (2006) to describe heat conduction for magnetized turbulent plasmas with $M_A<1$. In addition, Lazarian (2006) considered heat conduction by tubulence with $M_A>1$ as well as  heat advection by turbulence and compares the efficiencies of electron heat conduction and the heat transfer by turbulent motions.

~~Let us initially disregard the dynamics of fluid motions
on diffusion, i.e. consider diffusion induced by particles
moving along wandering turbulent magnetic field lines, which motions we disregard for the sake of simplicity.
Magnetized  turbulence with a dynamically important magnetic field is anisotropic with
eddies elongated along (henceforth denoted by $\|$) the direction of local magnetic field, i.e. $l_{\bot}<l_{\|}$, where $\bot$ denotes the
direction of perpendicular to the local magnetic field. Consider
isotropic injection of energy at the outer scale $L$ and
dissipation at the scale  $l_{\bot, min}$. This scale corresponds to the minimal dimension of the turbulent eddies.

~~Turbulence motions induce magnetic field divergence. It is easy to notice (LV99, NM01) that the separations of magnetic field lines at small scales less than the damping scale of turbulence, i.e. for $r_0<l_{\bot, min}$, are mostly influenced by
the motions at the smallest scale. This scale $l_{\bot, min}$ results in Lyapunov-type growth
$\sim r_0 \exp(l/l_{\|, min})$. This growth is
similar to that obtained in earlier models with a single scale of turbulent
motions (Rechester \& Rosenbluth 1978, henceforth RR78, Chandran \& Cowley 1998). Indeed, as the largest shear
that causes field line divergence
is due to the marginally damped motions at the scale around $l_{\bot, min}$ the effect of larger
eddies can be neglected and we are dealing with the case of single-scale "turbulence"
described by RR78.

~~The electron Larmor radius presents the minimal perpendicular scale of localization. Thus it is
natural to associate $r_0$ with the size of the cloud of electrons of the electron
Larmor radius $r_{Lar, particle}$. Applying the original RR78 theory (see also Chandran \& Cowley 1998)
they found that the electrons should travel over the distance
\begin{equation}
L_{RR}\sim l_{\|, min} \ln (l_{\bot, min}/r_{Lar, e})
\label{RR}
\end{equation}
 to get separated by $l_{\bot, min}$.

 ~~Within the single-scale "turbulent model" which formally corresponds to
$L{ss}=l_{\|, min}=l_{\bot, min}$ the distance $L_{RR}$
is called Rechester-Rosenbluth distance. For the ICM parameters the
logarithmic factor in Eq. (\ref{RR}) is
of the order of $30$, and this causes $30$ times decrease of thermal
conductivity for the single-scale models\footnote{For the single-scale model
$L_{RR}\sim 30L$ and the diffusion over distance $\Delta$ takes
$L_{RR}/L{ss}$ steps, i.e. $\Delta^2\sim L_{RR} L$, which decreases the
corresponding diffusion coefficient $\kappa_{e, single}\sim \Delta^2/\delta t$ by the factor of 30.}.

~~The single-scale "turbulent model" is just a toy model to study effects of turbulent motions. One can use this
model, however, to describe what is happening below the scale of the smallest eddies. Indeed, the shear and,
correspondingly, magnetic field line divergence is maximal for the marginally damped eddies at the dissipation
scale. Thus for scales less than the damping scale the action of the critically damped eddies is dominant.

~~In view of above, the realistic multi-scale turbulence
with a limited (e.g. a few decades) inertial range the single scale description is applicable for small scales up
to the damping scale. The logarithmic factor stays of the same order
but instead of the injection scale $L_{ss}$ for the single-scale RR model, one should use $l_{\|, min}$ for the actual
turbulence. Naturally, this addition does not affect the thermal conductivity,
provided that the actual turbulence injection scale $L$ is much larger than $\ l_{\|, min}$.
Indeed, for the electrons to diffuse isotropically they
should spread from $r_{Lar,e}$ to $L$. Alfvenic turbulence
 operates with field lines that are sufficiently stiff,
i.e. the deviation of the field lines from their original direction is
of the order unity at scale $L$ and less for smaller scales.
Therefore to get separated from the initial
distance of $l_{\bot, min}$ to a distance $L$ (see Eq. (\ref{subA}) with $M_A=1$), at which the motions get
uncorrelated, the electrons should diffuse the distance slightly larger (as
field lines are not straight) than
$\sqrt{2}L$. This is much larger than the extra
travel distance
$\sim 30 l_{\|, min}$ originating from sub-diffusive behavior at scales less than the turbulence damping scale. Explicit calculations in NM01 support this intuitive picture.

\subsection{Diffusion for $M_A>1$}

~~Turbulence with $M_A>1$ evolves along hydrodynamic
isotropic Kolmogorov cascade, i.e. $V_l\sim V_L (l/L)^{1/3}$ over the
range of scales $[L, l_A]$, where
\begin{equation}
l_A\approx  L (V_A/V_L)^{3}\equiv L M_A^{-3},
\label{lA}
\end{equation}
is the scale at which the magnetic field gets dynamically important, i.e.
 $V_l=V_A$. This scale plays the role
of the injection scale for the GS95 turbulence, i.e. $V_l\sim V_A (l_{\bot}/l_A)^{1/3}$,
with eddies at scales less than $l_A$
geting elongated in the direction of the local magnetic field.
The corresponding anisotropy can be characterized by the relation between
the semi-major axes of the eddies
\begin{equation}
l_{\|}\sim L (l_{\bot}/L)^{2/3} M_A^{-1},~~~ M_A>1,
\label{supA}
\end{equation}
 where
$\|$ and $\bot$ are related to the direction of the local magnetic field.
In other words, for $M_A>1$, the turbulence is still isotropic at the
scales larger to $l_A$, but
develops $(l_{\bot}/l_A)^{1/3}$ anisotropy for $l<l_A$.

~~If particles (e.g. electrons) mean free path
$\lambda\gg l_A$, they stream freely over the distance of $l_A$.
For particles initially at distance $l_{\bot, min}$ to get separated by $L$, the
required travel is the random walk with the step $l_A$, i.e. the mean-squared
displacement  of a particle till it enters an independent large-scale
eddy  $\Delta^2\sim l_A^2 (L/l_A)$, where $L/l_A$ is the number of steps.
These steps require time $\delta t\sim (L/l_A) l_A/C_1v_{e}$,
where $v_{particle}$ is electron thermal velocity and the coefficient $C_1=1/3$
accounts for 1D character of motion along magnetic field lines. Thus
the electron diffusion coefficient is
\begin{equation}
\kappa_{e}\equiv \Delta^2/\delta t\approx (1/3) l_A v_{e},~~~ l_A<\lambda,
\label{el}
\end{equation}
which for $l_A\ll \lambda$ constitutes a substantial reduction of diffusivity
compared to its unmagnetized value $\kappa_{unmagn}=\lambda v_{e}$.
We assumed in Eq. (\ref{el}) that $L\gg 30 l_{\|, min}$ (see \S 2.1).

~~For $\lambda\ll l_A\ll L$,  $\kappa_{e}\approx 1/3 \kappa_{unmagn}$ as both the $L_{RR}$
and the additional distance for electron to diffuse because of magnetic
field being stiff at scales less than $l_A$ are negligible compared
to $L$. For $l_A\rightarrow L$, when magnetic field has rigidity
up to the scale $L$, it gets around $1/5$ of
the value in unmagnetized medium, according to NM01.

\subsection{Diffusion for $M_A<1$}

 ~~It is intuitively clear that
for $M_A<1$ turbulence should be anisotropic from the injection scale $L$.
In fact, at large scales the turbulence is expected to be  {\it weak}\footnote{The terms ``weak'' and ``strong'' turbulence are accepted in the literature, but
can be confusing. As we discuss later at smaller scales at which the turbulent
velocities decrease the turbulence becomes {\it strong}. The formal theory of
weak turbulence is given in Galtier et al. (2000).}
(see Lazarian \& Vishniac 1999, henceforth LV99).
Weak turbulence is characterized
by wavepackets that do not change their $l_{\|}$, but develop structures
perpendicular to magnetic field, i.e. decrease $l_{\bot}$ . This cannot
proceed indefinitely, however. At some small scale
 the GS95 condition of {\it critical balance}, i.e. $l_{\|}/V_A\approx l_{\bot}/V_l$, becomes satisfied. This perpendicular scale $l_{trans}$
can be obtained substituting the scaling of
weak turbulence (see LV99) $V_l\sim V_L(l_{\bot}/L)^{1/2}$  into
the critical balance condition.
This provides $l_{trans}\sim L M_A^2$ and the corresponding
velocity $V_{trans}\sim V_L M_A$. For scales less than $l_{trans}$
the turbulence is {\it strong} and it follows the scalings of the
GS95-type, i.e. $V_l\sim V_L(L/l_{\bot})^{-1/3} M_A^{1/3}$ and
\begin{equation}
 l_{\|}\sim L (l_{\bot}/L)^{2/3} M_A^{-4/3},~~~ M_A<1.
\label{subA}
\end{equation}

~~For $M_A<1$, magnetic field wandering in the direction
perpendicular
to the mean magnetic field (along y-axis) can be described
by $d\langle  y^2\rangle/dx \sim \langle  y^2\rangle/l_\|$ (LV99),
where\footnote{The fact that
one gets $l_{\|, min}$ in Eq. (\ref{RR}) is related to the presence of this
scale in this diffusion equation.} $l_\|$ is expressed by Eq. (\ref{subA})
and one can associate $l_{\bot}$ with
$2\langle  y^2\rangle$
\begin{equation}
 \langle  y^2\rangle^{1/2} \sim \frac{x^{3/2}}{3^{3/2}L^{1/2}}
M_A^{2},~~~ l_{\bot}<l_{trans}
\label{3}
\end{equation}
For
weak turbulence $d\langle  y^2\rangle/dx \sim LM_A^4$ (LV99) and thus
\begin{equation}
\langle  y^2\rangle^{1/2} \sim L^{1/2} x^{1/2} M_A^2,~~~ l_{\bot}>l_{trans}.
\label{weak}
\end{equation}

\begin{figure*}
\centering
  \includegraphics[width=\textwidth]{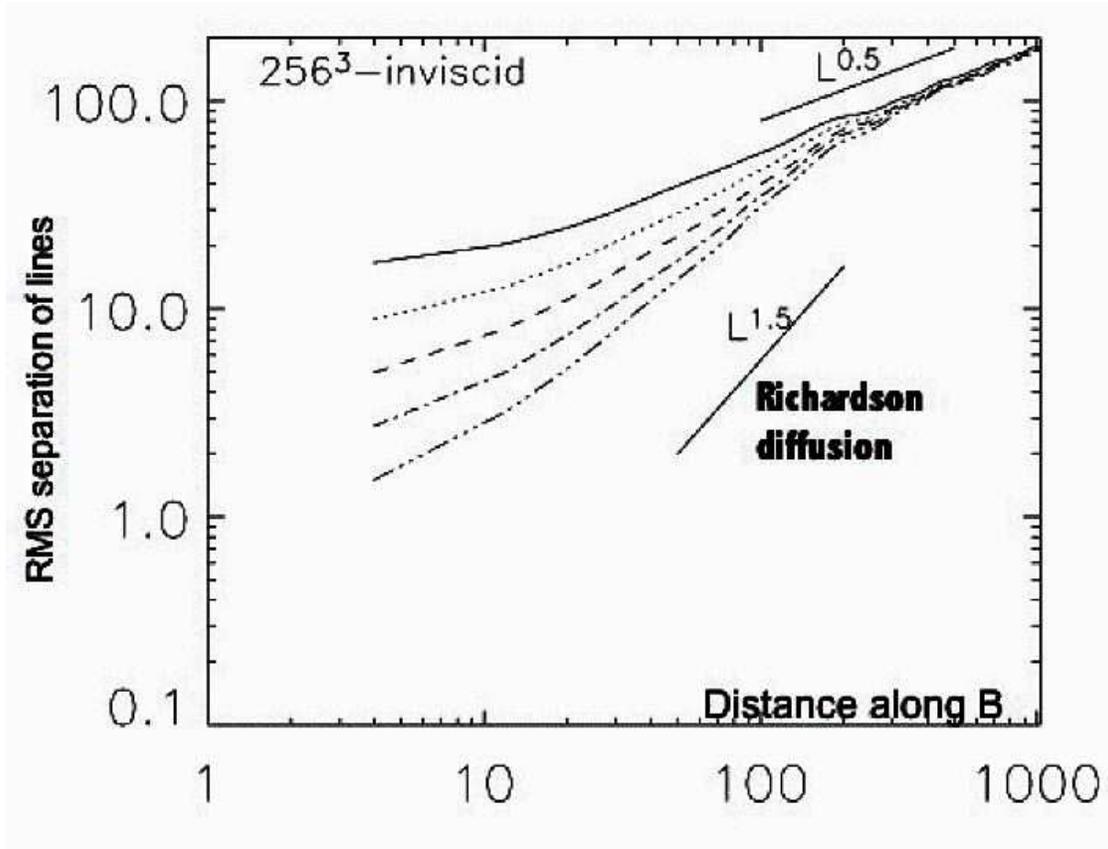}
\caption{Root mean square separation of field lines in a simulation of
inviscid MHD turbulence, as a function of
distance parallel to the mean magnetic field, for a range of initial
separations.  Each curve represents 1600 line pairs.
The simulation has been filtered to remove pseudo-Alfv\'en
modes, which introduce noise into the diffusion calculation. From Lazarian,
Vishniac \& Cho 2004.}
\label{LVC}
\end{figure*}

Fig.~\ref{LVC} confirms the correctness of the above scaling numerically.

~~Eq.~(\ref{3}) differs by the factor $M_A^{2}$ from that
in NM01, which reflects the gradual suppression of
thermal conductivity perpendicular to the mean magnetic field
as the magnetic field gets stronger. Physically this means that for
$M_A<1$ the magnetic field fluctuates around the well-defined
mean direction. Therefore the diffusivity gets anisotropic
with the diffusion coefficient parallel to the mean field
$\kappa_{\|, particle}\approx 1/3 \kappa_{unmagn}$ being larger than coefficient
for diffusion perpendicular to magnetic field $\kappa_{\bot, e}$.

~~Consider the coefficient $\kappa_{\bot, e}$ for $M_A\ll 1$. As
NM01 showed, particles become uncorrelated if they are displaced over the
 distance $L$ in the direction perpendicular to
 magnetic field.  To do this, a particle has first
to travel
$L_{RR}$ (see Eq.~(\ref{RR})),
 where Eq.~(\ref{subA}) relates $l_{\|, min}$ and $l_{\bot, min}$.
 Similar to the case in \S 2.1, for
$L\gg 30 l_{\|, min}$, the additional travel arising from the
 logarithmic factor is
negligible compared to the overall diffusion distance $L$.
 At larger scales electron has to diffuse
$\sim L$ in the direction parallel to magnetic field to cover the distance of
$L M_A^2$ in the direction
 perpendicular to magnetic field direction. To diffuse over a
distance R with random walk of
$LM_A^2$ one requires $R^2/L^2M_A^4$ steps. The time of the individual step
is $L^2/\kappa_{\|, e}$. Therefore the perpendicular diffusion
coefficient is
\begin{equation}
\kappa_{\bot, e}=R^2/(R^2/[\kappa_{\|, e}M_A^4])=\kappa_{\|, e}M_A^4,~~~ M_A<1,
\label{9}
\end{equation}
An essential assumption there is
that the particles do not trace their way back over the individual steps along
magnetic field lines, i.e. $L_{RR}<<L$. Note, that
for $M_A$ of the order of unity this is not accurate and
 one should account for the actual 3D displacement. This introduces the change
by a factor of order unity (see above).

\section{Transfer of heat through turbulent motions}

\begin{figure}
  \includegraphics[width=13.4cm]{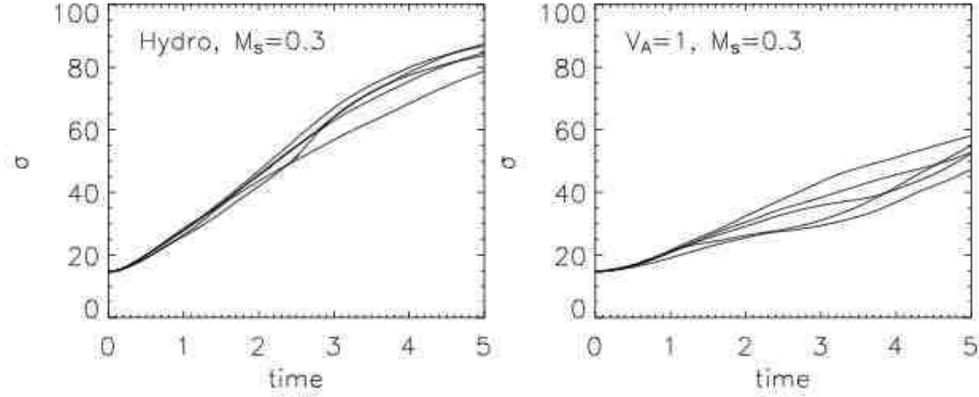}
  \caption{Comparison of the heat diffusion with time for hydro turbulence (left panel) and MHD transAlfvenic turbulence (right panel). Different curves correspond to different runs. From Cho et al. (2003).}
  \label{choetal}
\end{figure}

~~As we discussed above, turbulent motions themselves can induce advective transport of heat. Appealing to LV99 model of reconnection
one can conclude that turbulence
with $M_A\sim 1$ should be similar to hydrodynamic turbulence, i.e.
\begin{equation}
\kappa_{dynamic}\approx C_{dyn} L V_L,~~~ M_A>1,
\label{dyn}
\end{equation}
where $C_{dyn}\sim 0(1)$ is a constant, which for hydro turbulence
is around $1/3$ (Lesieur 1990). This was confirmed in Cho et al. (2003) (see Figure \ref{choetal} and also Cho \& Lazarian 2004) where
MHD calculations were performed for transAlfvenic turbulence with $M_A\sim 1$. As large scale eddies of superAlfvenic turbulence are essentially hydrodynamic, the correspondence between the ordinary hydrodynamic heat advection and superAlfvenic one should only increase as $M_A$ increases.

~~If we deal with heat transport, for fully ionized non-degenerate plasmas we assume $C_{dyn}\approx 2/3$ to account for the advective heat transport by
both protons and electrons\footnote{This becomes clear if one uses the
heat flux equation
$q=-\kappa_c\bigtriangledown T$, where $\kappa_c=nk_B\kappa_{dynamic/electr}$,
$n$ is {\it electron} number density, and $k_B$ is the Boltzmann constant,
for both electron and advective heat transport.}.
Thus eq.~(\ref{dyn}) covers the cases of
both $M_A>1$ up to $M_A\sim 1$. For $M_A<1$
one can estimate $\kappa_{dynamic}\sim d^2\omega$, where $d$ is
the random walk of the field line over the wave period $\sim \omega^{-1}$.
As the weak turbulence
at scale $L$ evolves over time $\tau\sim
M_A^{-2}\omega^{-1}$, $\langle y^2 \rangle$  is the result of
the random walk with a step $d$, i.e.
$\langle y^2 \rangle\sim (\tau\omega)d^2$. According to eq.(\ref{3}) and
(\ref{weak}), the  field line is displaced over  time $\tau$ by
$\langle y^2\rangle\sim L M_A^4 V_A \tau$. Combining the two one gets
 $d^2 \sim L M_A^3 V_L \omega^{-1}$, which provides
$\kappa_{dynamic}^{weak}\approx C_{dyn} LV_L M_A^3$,
which is similar to the diffusivity arising from strong turbulence at
scales less than $l_{trans}$, i.e. $\kappa_{dynamic}^{strong}\approx C_{dyn} l_{trans}
V_{trans}$. The total diffusivity is the sum of the two, i.e. for
plasma
\begin{equation}
\kappa_{dynamic}\approx (\beta/3) LV_L M_A^3, ~~~ M_A<1,
\label{dyn_weak}
\end{equation}
where $\beta\approx 4$.

\section{Relative importance of two processes}

\subsection{General treatment}

\begin{figure}
  \includegraphics[width=13.8cm]{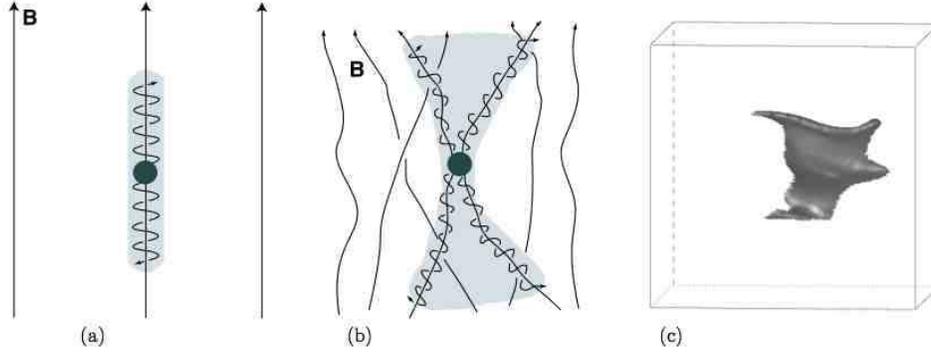}
  \caption{(a) The textbook description of confinement of charged particles in magnetic fields; (b) diffusion of particles in turbulent fields; (c) advection of heat from a localized souce by eddies in MHD numerical simulations. From Cho \& Lazarian 2004.}
  \label{cho}
\end{figure}

~~Figure \ref{cho} illustrates the existing ideas on processes of heat conduction in astrophysical plasmas. They range from the heat insulation by unrealistically laminar magnetic field (see panel (a)), to heat diffusion in turbulent magnetic field (see panel (b)) and to heat advection by turbulent flows (see panel (c)). The relative efficiencies of the two latter processes depend on parameters of turbulent plasma.

\begin{figure}
\centering
  \includegraphics[height=.45\textheight]{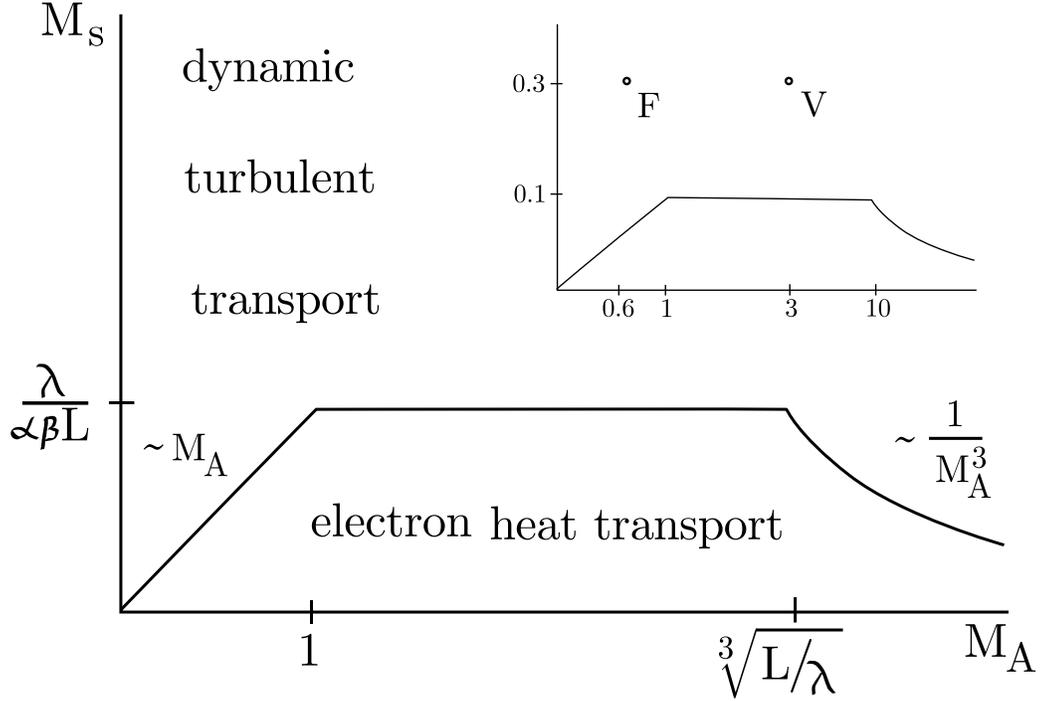}
  \caption{Parameter space for particle diffusion or turbulent diffusion to
dominate: application to heat transfer. Sonic Mach number $M_s$ is ploted
against the Alfven Mach number $M_A$. The heat transport is dominated by the
dynamics of turbulent eddies is above the curve (area denoted "dynamic turbulent
transport") and by thermal conductivity of
 electrons is below the curve (area denoted "electron heat transport"). Here $\lambda$ is the mean free path of the
electron, $L$ is the driving scale, and $\alpha=(m_e/m_p)^{1/2}$, $\beta\approx 4$. {\it Example of theory application}: The panel in the right upper corner of the figure illustrates
heat transport for the parameters for a cool
core Hydra cluster (point ``F''), ``V'' corresponds
to the illustrative model of a cluster core in Ensslin et al. (2005).  Relevant parameters were used 
for $L$ and $\lambda$. From
Lazarian (2006).}
\label{space}
\end{figure}

~~In thermal plasma, electrons are mostly responsible for thermal conductivity.
The schematic of the parameter space for  $\kappa_{particle}<\kappa_{dynamic}$ is shown in Fig~\ref{space}, where the
the Mach number $M_s$
 and the Alfven Mach number $M_A$ are
the variables. For $M_A<1$, the ratio of diffusivities arising from
fluid and particle motions is
 $\kappa_{dynamic}/\kappa_{particle}\sim \beta\alpha M_S M_A(L/\lambda)$ (see
Eqs. (\ref{9}) and (\ref{dyn_weak})), the square root of the
ratio of the electron to proton mass $\alpha=(m_e/m_p)^{1/2}$, which
provides the separation line between the two regions in Fig. 2, $\beta\alpha M_s\sim
(\lambda/L) M_A$. For $1<M_A<(L/\lambda)^{1/3}$ the mean free path is less
than $l_A$ which results in $\kappa_{particle}$ being some fraction of
$\kappa_{unmagn}$, while $\kappa_{dynamic}$ is  given by Eq.~(\ref{dyn}).
Thus $\kappa_{dynamic}/\kappa_{particle}\sim \beta \alpha M_s (L/\lambda)$, i.e. the ratio
 does not depend on $M_A$ (horisontal line in Fig.~2). When $M_A>(L/\lambda)^{1/3}$ the
mean free path of electrons is constrained by $l_A$. In
this case $\kappa_{dynamic}/\kappa_{particle}\sim \beta \alpha M_s M_A^3$ (see
Eqs. (\ref{dyn})
and (\ref{el})) . This results in the separation line
$\beta\alpha M_s \sim M_A^{-3}$
in Fig.~\ref{space}.

\subsection{Application to ICM plasmas}

~~Consider plasmas in clusters of galaxies to
illustrate the relative importance of two
processes of heat transfer. Below we shall provide evidence that {\it magnetized} Intracluster Medium (ICM) is turbulent and therefore our considerations above should be applicable. 

~~It is generally believed that ICM plasma is turbulent. However, naive estimates of diffusivity
for collisionless plasma provide numbers which may cast doubt on this conclusion. Indeed, in unmagnatized
plasma with the ICM temperatures $T\sim 10^8$ K and and density $10^{-3}$
cm$^{-3}$ the kinematic viscosity $\eta_{unmagn} \sim v_{ion} \lambda_{ion}$,
where $v_{ion}$ and $\lambda_{ion}$ are the velocity of an ion and
its  mean free path, respectively, would make the Reynolds number
$Re\equiv LV_L/\eta_{unmagn}$ of the order of 30. This is barely enough for the onset
of turbulence. For the sake of simplicity we assume that ion mean free path coincides with the proton mean free path and both scale as $\lambda\approx
3 T_3^2 n_{-3}^{-1}$~kpc, where the temperature $T_3\equiv kT/3~{\rm keV}$ and
$n_{-3}\equiv n/10^{-3}~{\rm cm^{-3}}$. This provides
$\lambda$  of the order of 0.8--1 kpc for
the ICM (see NM01). We shall argue that the above low estimate of $Re$ is an artifact of our 
neglecting magnetic field. 

~~ In general, a single value of $Re$ uniquely characterizes hydrodynamic flows. The case of magnetized
plasma is very different as the diffusivities of protons parallel and perpendicular to magnetic fields are
different.  The diffusion of protons perpendicular to the local magnetic field is usually very slow. Such a 
diffusion arises from proton scattering. Assuming the maximal
scattering rate of an proton, i.e. scattering every orbit (the
so-called Bohm diffusion limit) one gets the viscosity perpendicular
to magnetic field $\eta_{\bot}\sim v_{ion} r_{Lar, ion}$, which is much smaller
than $\eta_{unmagn}$, provided that the ion Larmor radius $r_{Lar, ion}\ll \lambda_{ion}$.
For the parameters
of the ICM this allows essentially inviscid fluid motions\footnote{A {\it regular} magnetic field
 $B_\lambda\approx (2 m k T)^{1/2} c/(e\lambda)$ that makes
$r_{Lar, ion}$ less than $\lambda$ and therefore $\eta_{\bot}<\nu_{unmagn}$
is just $10^{-20}$~G. {\it Turbulent} magnetic field with many reversals
over $r_{Lar, ion}$ does not interact efficiently with a proton, however. As the
result, the protons are not constrained
until $l_A$ gets of the order of $r_{Lar, ion}$. This happens when
the turbulent magnetic field is of the
 order of $2\times 10^{-9}(V_L/10^3{\rm km/s})$~G. At this point, the step for
the random walk is $\sim 2\times 10^{-6}$~pc and the Reynolds
number is $5\times 10^{10}$. }
 of magnetic lines
parallel to each other, e.g. Alfven motions.

~~In spite of the substantial progress in understading  of the ICM
 (see En{\ss}lin, Vogt \& Pfrommer 2005, henceforth EVP05, En{\ss}lin \&
Vogt 2006, henceforth EV06
and references therein), the basic parameters of ICM
turbulence are known within the factor of 3 at best. For instance, the
estimates of  injection velocity $V_L$ varies in the literature
from 300 km/s to $10^3$ km/s,
while the injection scale $L$ varies from 20 kpc to 200 kpc,
depending whether
the injection of energy by galaxy mergers or galaxy wakes is considered.
EVP05 considers an {\it illustrative} model in which the magnetic field
 with the
10 $\mu$G fills 10\% of
the volume, while 90\% of the volume is filled with the field of $B\sim 1$
 $\mu$G. Using the latter number and  assuming
 $V_L=10^3$ km/s, $L=100$ kpc, and the density of the hot ICM
is $10^{-3}$ cm$^{-3}$, one gets $V_A\approx 70$ km/s, i.e. $M_A>1$.
 Using the numbers above, one
 gets $l_A\approx 30$ pc for the 90\% of the volume of the hot ICM,
which is much less than $\lambda_{ion}$. The diffusivity of ICM plasma
gets $\eta=v_{ion} l_A$ which for the parameters above provides
 $Re\sim 2\times 10^3$, which is enough for driving superAlfvenic turbulence
at the outer scale $L$. However, as $l_A$ increases as $\propto B^3$, $Re$ gets around $50$ for
the  field of 4 $\mu$G, which
is at the border line of exciting turbulence\footnote{One can imagine
dynamo action in which superAlfvenic turbulence generates magnetic
field till $l_A$ gets large enough to shut down the turbulence.
}. However, the regions with higher
magnetic fields (e.g. 10 $\mu$G)
can support Alfvenic-type turbulence with the injection scale $l_A$ and
the injection velocities resulting from large-scale shear
$V_L (l_A/L)\sim V_L M_A^{-3}$.

~~For the regions of $B\sim 1$ $\mu$G the value of $l_A$ is
smaller than the mean free path of electrons $\lambda$.
 According to Eq.~(\ref{el}) the value of
$\kappa_{electr}$ is 100 times smaller than $\kappa_{Spitzer}$.
On the contrary, $\kappa_{dynamic}$ for the ICM parameters adopted will be
$\sim 30 \kappa_{Spitzer}$, which makes
 the heat transfer by turbulent motions the dominant process. This agrees well
with the observations in Voigt \& Fabian (2004). Fig.~2 shows the dominance
of advective heat transfer for the parameters
of the cool core of Hydra A ( $B=6$~$\mu$G,
$n=0.056$ cm$^{-3}$, $L=40$~kpc, $T=2.7$~keV according to
EV06), point ``F'', and for the illustrative model in EVP05, point ``V'',
for which $B=1$~$\mu$G (see also Lazarian 2006).

~~Note that our stationary model
of MHD turbulence is not directly applicable to transient
wakes behind galaxies.
 The ratio of the
damping times of the hydro turbulence and the time of
 straightening of the magnetic field
lines is $\sim M_A^{-1}$. Thus, for $M_A>1$, the magnetic field at scales
larger than $l_A$ will be straightening
gradually after the hydro turbulence has faded away over time $L/V_L$.
The process can be characterized as injection of turbulence at velocity
$V_A$ but at scales that increase linearly with time, i.e. as $l_A+V_At$.
 The study of heat transfer in transient turbulence and
 magnetic field ``regularly'' stretched by passing galaxies
 is an interesting process that requires further investigation.

 \section{Richardson diffusion and SuperDiffusion on small scales}

 ~~All the discussion above assumed that we deal with diffusion within magnetized plasmas over
 the scales much larger than the turbulence injection scale $L$. Below we show that on
 the scales less than $L$ we deal with non-stationary processes.

 \subsection{Richardson-type advection of heat}

~~The advection of heat on scales less than the turbulent injection scale $L$ happens through smaller scale eddies. Thus the earlier
 estimate of turbulent diffusion of heat in terms of the injection velocity and the injection scale does not apply. In the lab system of reference the transfer of heat is difficult to describe and one should use the Lagrangian description.

~~One can consider two-particle turbulent diffusion or Richardson diffusion by dealing with the separation
$\Bell(t)=\bx(t)-\bx'(t)$ between a pair of Lagrangian fluid particles (see Eyink et al. 2011). It was proposed
by Richardson (1926) that this separation grows in turbulent flow according to the
formula
\be \frac{d}{dt}\langle \ell_i(t)\ell_j(t)\rangle=\langle
\kappa_{dynanic, ij}(\Bell)\rangle \lb{richardson} \ee
with a scale-dependent eddy-diffusivity $\kappa_{dynamic}(\ell).$ In hydrodynamic turbulence
Richardson deduced that $\kappa_{dynamic}(\ell)\sim \varepsilon^{1/3}\ell^{4/3}$ (see Obukhov 1941) and thus $\ell^2(t)\sim \varepsilon t^3. $
 An analytical
formula for the 2-particle eddy-diffusivity was derived by Batchelor (1950) and Kraichnan (1966):
\begin{equation}
  \kappa_{dynamic, ij}(\Bell)=\int_{-\infty}^0 dt \langle \delta U_i(\Bell,0)
\delta U_j(\Bell,t)\rangle
  \end{equation}
with $ \delta U_i(\Bell,t)\equiv U_i(\bx+\Bell,t)-U_i(\bx,t)$ the relative velocity
at time $t$ of a pair of fluid particles which were at positions $\bx$ and $\bx+\Bell$
at time 0.

~~How can one understand these results? Consider a hot spot of the size $l$ in a turbulent flow. The spot is going to be mostly expanded by turbulent eddies of size $l$. The turbulent velocity $u(l)=\frac{d}{dt} l(t)$ for Kolmogorov turbulence is proportional
to $l^{1/3}$. Performing formal integration one gets an asymptotic solution for large time scales $l^2(t)\sim t^3$, which corresponds
to the Richardson diffusion law. Physically, as the hot spot extends, it is getting sheared by larger and eddies, which induce the
accelerated expansion of the hot spot.

~~For magnetic turbulence the Kolmogorov-like description is valid for motions induced by strong Alfvenic turbulence in the direction
perpendicular to the direction of the local magnetic field\footnote{The local magnetic field direction fluctuates in the lab system of reference. Thus the results of the diffusion in the lab system are less anisotropic.}. Thus we expect that Richardson diffusion to be applicable to the magnetized turbulence case.

 \subsection{Superdiffusion of heat perpendicular to mean magnetic field}

~~The effects related to the diffusion of heat via electron streaming along magnetic field lines are different when the problem is
 considered at scales $\gg L$ and  $\ll L$. This difference is easy to understand as on small scales magnetized eddies are very elongated, which means that the magnetic field lines are nearly parallel. However, as electrons diffuse into larger eddies, the
dispersion of the magnetic field lines in these eddies gets bigger and the diffusion perpendicular to the mean magnetic field increases\footnote{Below we consider turbulent scales that are larger than the electron mean free path $\lambda_e$. Heat transfer at smaller scale is not a diffusive process, but happens at the maximal rate determined by the particle flux $n v_{th}$ provided that
we deal with scales smaller than $l_A$. The perpendicular to magnetic field flux is determined by the field line deviations on the given scale as we discussed above (see also
LV99).}

~~{\it SuperAlfvenic turbulence}:\\
~~On scales $k_\parallel^{-1}<l_A$, i.e., on scales at which magnetic fields are strong enough to influence
turbulent motions,
the mean deviation of a field in a distance $k_\|^{-1}=\delta z $ is given by LV99 as
\be
<(\delta x)^2>^{1/2}=\frac{([\delta z] M_A)^{3/2}}{3^{3/2}L^{1/2}},~~~M_A>1
\label{highmwand}
\ee
Thus, for scales much less than $L$ (see also Yan \& Lazarian 2008)
\be
\kappa_{e, \bot}\approx
\left(\frac{\delta x}{\delta z}\right)^2 \kappa_{e,\|}\sim \frac{[\delta z] M_A^3}{3^3L} \kappa_{e,\|}\sim \kappa_\| (k_\|l_A)^{-1}, ~~~M_A>1,
\label{dw1}
\ee
which illustrates the non-stationary regime of {\it superdiffusion}, where the diffusion coefficient changes with the scale $k_{e,\|}^{-1}$.

~~{\it SubAlfvenic turbulence}:\\
~~On scales larger than $l_{tr}$, the turbulence is weak. The mean
 deviation of a field in a distance $\delta z$ is given by Lazarian (2006):
\be
<(\delta x)^2>^{1/2}=\frac{[\delta z]^{3/2}}{3^{3/2}L^{1/2}}M_A^2,~~~M_A<1.
\label{lowmwand}
\ee

~~For the scales $L>k_\parallel^{-1}=\delta z$
we combine Eq.~(\ref{lowmwand}) with
\be
\delta z=\sqrt{kappa_{e,\|} \delta t}
\label{zz}
\ee
 and get for scales much less than $L$
\be
\kappa_{e,\bot}\approx
\frac{\delta x^2}{\delta t}=\frac{\kappa_{e,\|}\delta z}{3^3L}M_A^4\sim \kappa_{e,\|}(k_\|L)^{-1}
M_A^4,
\label{dw2}
\ee
which for a limiting case of $k_{e, \|}\sim L^{-1}$ coincides up to a factor
with the Eq.~(\ref{9}).

~~Eqs.~(\ref{dw1}) and (\ref{dw2}) certify that the perpendicular diffusion at scales much less than the injection scale accelerates as z grows.

\subsection{Comparison of processes}

~~Both processes of heat transport at the scales less than the turbulence injection scale are different from the diffusion at large scales as the rate of transport depends on the scale. However, the description of heat transport by electrons is more related to the measurements in the lab system. This follows from the fact that the dynamics of magnetic field lines is not important for the process and it is electrons which stream along wandering magnetic field lines. Each of these wandering magnetic field lines are snapshot of the magnetic field line dynamics as it changes through magnetic reconnection its connectivity in the ambient plasma. Therefore the description of heat transfer is well connected to the lab system of reference. On the contrary, the advection of heat through the Richardson diffusion is a process that is related to the Lagrangian description of the fluid. Due to this difference the direct comparison of the efficiency of processes is not so straightforward.

~~For example, if one introduces a localized hot spot, electron transport would produce heating of the adjacent material along the expanding cone of magnetic field lines, while the turbulent advection would not only spread the hot spot, but also advect it by
the action of the largest eddies.

 \section{Outlook on the consequences}

~~Magnetic thermal insulation is a very popular concept in astrophysical literature dealing with magnetized plasmas. Our discussion above shows that in many cases this insulation is very leaky. This happens due to ubiquitous astrophysical turbulence which induces magnetic field wandering and interchange of pieces of magnetized
 plasma enabled by turbulent motions. Both processes are very closely related to the process of fast magnetic
 reconnection of turbulent magnetic field (LV99).

~~As a result, instead of an impenetrable wall of laminar ordered magnetic field lines, the actual turbulent field lines present a complex network of tunnels along which electrons can carry heat. As a result, the decrease of heat conduction amounts to a factor in the range of 1/3 for mildly superAlfvenic turbulence to a factor $\sim 1/5$ for transAlfvenic turbulence. The cases when heat conductivity by electrons may be suppressed to much greater degree include highly superAlfvenic turbulence and highly subAlfvenic turbulence. In addition, turbulent motions
induce heat advection which is similar to turbulent diffusivity of unmagnetized fluids.

 ~~The importance of magnetic reconnection cannot be stressed enough in relation to the process of heat
 transfer in magnetized plasmas. As a consequence of fast magnetic reconnection plasma does not stay entrained on the same magnetic field lines, as it is usually presented in textbooks. On the contrary, magnetic field lines constantly change their connectivity and plasma constantly samples newly formed magnetic field lines enabling efficient diffusion. Therefore we claim that the advection of heat by turbulence is an example of a more general process of reconnection diffusion. It can be noticed parenthetically that the turbulent advection of heat is a well knows process. However, for decades the discussion of the process avoided in astrophysical literature due the worries of the effect of reconnection that inevitably should accompany it. The situation has changed with better understanding of magnetic reconnection in turbulent environments (LV99). It worth pointing out that our estimates indicate that in many astrophysicaly important cases, e.g. for ICM, the advective heat transport by dynamic turbulent eddies dominates thermal conductivity.

~~Having the above processes in hand, one can describe heat transport within magnetized astrophysical plasmas.
For instance, we discussed the heat transfer by
particle and turbulent motions for $M_A<1$ and $M_A>1$.
It is important that we find that turbulence can both enhance diffusion
 and suppress it. We showed that when
$\lambda$ gets larger than $l_A$ the
conductivity of the medium $\sim M_A^{-3}$ and therefore the turbulence
{\it inhibits} heat transfer, provided that
$\kappa_{e}>\kappa_{dynamic}$. Along with the plasma effects that we mention
below, this effect can, indeed, support
sharp temperature gradients in  hot plasmas
with weak magnetic  field.

~~As discussed above, rarefied plasma, e.g. ICM plasma,
 has large viscosity for motions parallel
to magnetic field and marginal viscosity for motions that induce
perpendicular mixing. Thus fast dissipation of sound waves in the ICM
does not contradict the medium being turbulent. The later may be
important for the heating of central regions of clusters caused by
 the AGN feedback
(see Churasov et al. 2001, Nusser, Silk \& Babul 2006 and more
references in EV06). Note, that
 models that include both heat transfer from the outer hot regions and an additional heating from the AGN feedback
look rather promissing (see Ruszkowkski \& Begelman 2002,
Piffaretti \& Kaastra 2006).
We predict that the viscosity for
1 $\mu$G regions is
less than for 10 $\mu$G regions and therefore heating by sound waves (see Fabian
et al. 2005) could be more efficient for the latter.
Note, that the plasma instabilities in collisionless magnetized ICM
arising from compressive motions (see Schekochihin \& Cowley 2006,
Lazarian \& Beresnyak 2006) can resonantly scatter particles
and decrease $\lambda$.
This decreases further $\kappa_{e}$ compared to $\kappa_{unmagn}$ but
increases $Re$.
 In addition, we disregarded mirror effects that can reflect electrons
back\footnote{Many of these papers do not use realistic models of turbulence and therefore overestimate the effect of electron reflection.} (see Malyshkin \& Kulsrud 2001 and references therein),
which can further decrease $\kappa_{e}$. While there are many instabilities that are described in plasmas with temperature gradient, many of those are of academic interest, as they do not take into account the existence of ambient turbulence.

~~For years the attempts to describe heat transfer in magnetized plasma were focused on
finding the magic number which would be the reduction factor characterizing the effect of magnetic
field on plasmas' diffusivity. Our study reveals a different and more complex picture.
 The heat transfer depends on sonic and
Alfven Mach numbers of turbulence and the corresponding diffusion coefficient
vary substantially for plasmas with different level of magnetization and turbulent excitation.
In different astrophysical environments turbulence can both inhibit or enhance diffusivity depending
on the plasma magnetization and turbulence driving.

~~The issues of ``subdiffusivity'' or magnetic field retracing their paths was a worrisome issue that for years impeded
the progress in understanding heat transport in plasmas. We claim that the retracing does happen, but on the scales which are of the order of the eddies at the dissipation scale. As an electron has a finite Larmor radius in the retracing the same magnetic field line it experiences the deviations from its original trajectory. On the scale less than the dissipation scale these deviations grow from the electron Larmor radius in accordance with Lyapunov exponents, but on larger scale the separation is determined by field wandering only and does not depend on the Larmor radius. Thus the effect of retracing for heat transfer in real-world astrophysical turbulence with a substantial separation of the turbulence injection scale and dissipation scales is marginal.

~~On the contrary, the issue of "superdiffusivity" may be important for heat transfer on the scales less than the turbulence injection scale. Richardson diffusion or more correctly its anisotropic analog present in magnetized plasma (see Eyink et al. 2011) is an example of superdiffusion induced by eddies of increasing size. A similar effect is also true for magnetic field line wandering. The effect of "superdiffusive" heat transfer requires additional studies.

~~It is worth mentioning that another parameter that determines the heat flux into the magnetized volume is the area
of the contact of plasmas with different temperatures. For instance, if the magnetic flux is "shredded", i.e. consists of numerous separated individual flux tubes, then the heating of plasma within magnetized tubes may be more efficient. For instance, Fabian et al. (2011) appealed to reconnection diffusion of ambient plasma into "shredded" magnetic flux of NGC1275 in Perseus cluster in order to explain heating and ionization of the magnetic filaments.

~~In view of the discussion above one can conclude that realistically turbulent magnetic fields do not completely suppress heat conductivity of astrophysical plasmas. The decrease of thermal conductivity depends on the Alfven Mach number of turbulence. At the same time, turbulent motions enhance heat transport via heat advection. In special situations, e.g. in very weakly turbulent magnetic field, the transport of heat in plasmas may still be slow.


{\bf Acknowledgments}  The research is supported by
 the NSF grant AST 0808118 and the Center for Magnetic Self Organization in
Laboratory and Astrophysical Plasmas (CMSO).


\end{document}